\documentclass[pdflatex,sn-nature]{sn-jnl}

\usepackage{soul}
\usepackage{graphicx}%
\usepackage{multirow}%
\usepackage{amsmath,amssymb,amsfonts}%
\usepackage{amsthm}%
\usepackage{mathrsfs}%
\usepackage[title]{appendix}%
\usepackage{xcolor}%
\usepackage{textcomp}%
\usepackage{manyfoot}%
\usepackage{booktabs}%
\usepackage{algorithm}%
\usepackage{algorithmicx}%
\usepackage{algpseudocode}%
\usepackage{listings}%

\theoremstyle{thmstyleone}%
\theoremstyle{thmstyletwo}%

\theoremstyle{thmstylethree}%

\begin{document}

\title[Article Title]{Single-exposure holographic lithography of ultra-high aspect-ratio microstructures}


\author[1]{\fnm{Dajun} \sur{Lin}}

\author[2]{\fnm{Brian} \sur{Baker}}

\author[1]{\fnm{Rajesh} \sur{Menon}*}\email{*rmenon@eng.utah.edu}

\affil[1]{\orgdiv{Department of Electrical \& Computer Engineering}, \orgname{University of Utah}, \orgaddress{\street{50 Central Campus Dr.}, \city{Salt Lake City}, \postcode{84112}, \state{UT}, \country{USA}}}

\affil[2]{\orgdiv{Utah nanofab}, \orgname{University of Utah}, \orgaddress{\street{36 S. Wasatch Drive}, \city{Salt Lake City}, \postcode{84112}, \state{UT}, \country{USA}}}

\abstract{Volumetric lithography offers a path to scalable fabrication of complex three-dimensional (3D) micro- and nanoscale architectures, yet existing approaches are limited by quasi-two-dimensional exposure physics or slow serial writing. We present a single-exposure volumetric fabrication strategy that enables creation of ultrahigh-aspect-ratio 3D structures with 6~\textmu m minimum features. An inverse-designed volumetric (holographic) phase mask generates an extended-depth-of-field intensity distribution inside a photoresist volume while preserving high transverse resolution, enabling uniform polymerization of the full volume in a single exposure. With exposure times of approximately 20~s, we fabricate lattices, Penrose tilings, and micromechanical elements with feature sizes down to 6~\textmu m over volumes up to 800 × 800 × 720~\textmu m\textsuperscript{3}, achieving aspect ratios exceeding 120:1. Quantitative analysis of capillary flow in hollow lattices demonstrates controlled fluid transport with an effective capillary transport coefficient of 176.3~\textmu m/$\sqrt{\text{ms}}$. In situ nanoindentation-based micro-compression reveals that the printed 3D hexagonal close-packed lattices exhibit a well-defined linear elastic regime with an effective Young’s modulus of 5.7~GPa, followed by progressive buckling and densification characteristic of mechanically robust cellular architectures. Overlapping, tilted and multi-mask exposures further enable quasi-3D complex geometries with potential for reconfigurability. This approach establishes a new regime of high-throughput volumetric fabrication.}




\maketitle

\section{Introduction}

Three-dimensional micro- and nanofabrication\cite{surjadi2025enabling, bernal2025road,gu20253d,skliutas2025multiphoton,somers2024physics} underpins a broad range of emerging technologies, including micro-optical systems, metasurfaces,\cite{hu2025height} micro-electromechanical systems (MEMS),\cite{miyajima1995high} microfluidics, soft robotics,\cite{leber2023highly} and architected materials \cite{sun2019twophoton}. Many of these applications require structures that simultaneously combine large fabrication volumes, sub--10-\textmu m lateral features, and ultra-high aspect ratios. Achieving these attributes at fabrication speeds\cite{hahn2020rapid} compatible with scalable manufacturing remains a central challenge in micro- and nanoscale fabrication.

Conventional photolithography is fundamentally optimized for planar patterning and is therefore intrinsically limited in achievable thickness. While grayscale lithography and multi-exposure stacking approaches can produce limited 3D relief structures, they rely on serial layer accumulation and suffer from interlayer stitching errors, accumulated alignment tolerances, and degraded mechanical integrity in tall structures. As a result, fabrication time increases rapidly with structure height, and true volumetric patterning remains impractical.

Serial point-writing techniques, most notably two-photon polymerization (2PP),\cite{obata2013high} enable true 3D fabrication with submicrometer resolution and arbitrary geometry \cite{maruo1997two}. However, 2PP is constrained by voxel-by-voxel writing speeds that typically range from $10^{3}$ to $10^{4}$ voxels s$^{-1}$, even with optimized scanning strategies \cite{sun2019twophoton}. Consequently, fabrication of millimetre-scale volumes often requires hours, limiting throughput and scalability despite advances in parallelization and scan optimization.

To overcome these limitations, volumetric additive manufacturing (VAM) methods have emerged as a promising alternative. Techniques such as computed axial lithography \cite{toombs2022volumetric, kelly2019volumetric, debeer2019cal}, xolography \cite{regehly2020xolography}, and related multi-beam volumetric exposure schemes \cite{bernardo2022vam} enable layerless fabrication by distributing optical dose throughout a resin volume using angular or spectral multiplexing. These approaches dramatically reduce fabrication time, often achieving full part formation within seconds to minutes. However, current VAM implementations are fundamentally limited in spatial resolution, typically to feature sizes of 25--100~\textmu m, due to constraints imposed by resin diffusion,\cite{orth2023deconvolution} oxygen inhibition, and nonlinear exposure thresholds \cite{lin2026single, bernardo2022vam}. These limitations preclude their use for micro- and nanostructured devices requiring fine features and high aspect ratios.

Single-exposure photolithography provides an attractive pathway to mechanically robust, high-aspect-ratio structures by eliminating layer interfaces altogether. Traditional hard- and soft-contact mask lithography can generate elongated axial intensity profiles, but diffraction-induced proximity effects rapidly degrade lateral resolution with increasing propagation distance. As a result, usable fabrication depths are typically limited to tens of micrometres, and feature merging becomes unavoidable for thick photoresist layers. Analytical structured-light approaches have demonstrated single-shot fabrication of simple 3D geometries, including helices and tubular structures, but these methods are generally restricted to certain geometries and thus far have not demonstrated extremely large aspect ratios.

Diffraction-invariant and extended-depth-of-field optical fields have been extensively explored in imaging,\cite{banerji2020extreme, menon2023perspectives, hayward2024imaging} wavefront engineering,\cite{jena2025extended, kumari2024generating} and holography,\cite{lin2025diffraction}. These concepts suggest that carefully engineered optical fields can maintain lateral confinement over extended axial ranges, analogous to nondiffracting beams or extended-depth-of-focus imaging systems. However, translating these ideas into practical volumetric photolithography with high spatial fidelity, arbitrary geometries, and scalable fabrication speed has remained an open challenge.

\begin{figure}[h]
\centering
\includegraphics[width=1\textwidth]{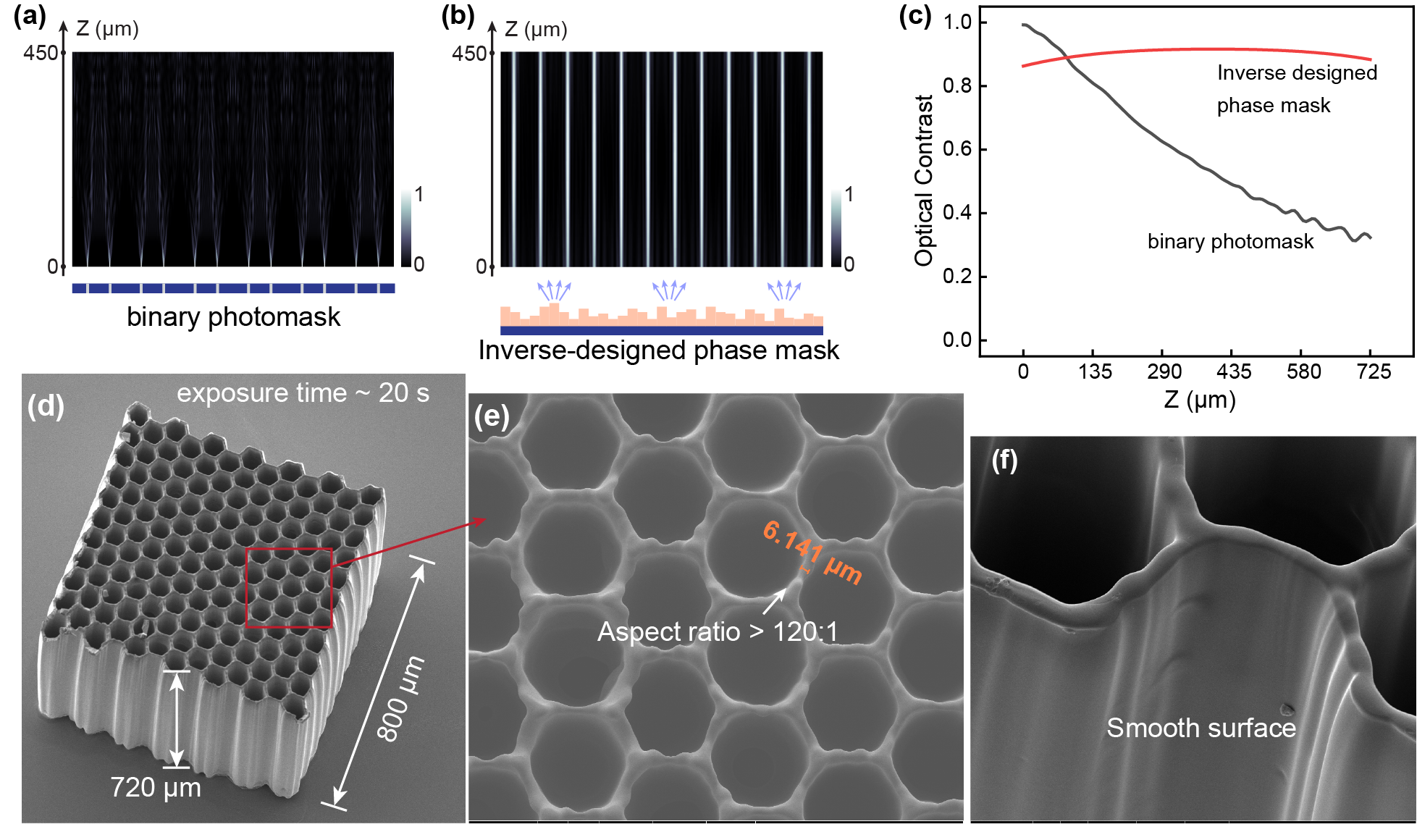}
\caption{\textbf{Single-exposure fabrication of $>$120:1 aspect-ratio microstructures.} (a) A conventional binary photomask produces only low-aspect-ratio features because diffraction rapidly degrades optical contrast during propagation. (b) An inverse-designed phase mask suppresses diffraction \cite{menon2023perspectives}, enabling the projection of high-aspect-ratio geometries deep into a photoresist film, and allowing true single-exposure volumetric fabrication. (c) Simulated optical contrast as a function of propagation distance from the mask shows that the binary mask loses contrast rapidly, whereas the inverse-designed phase mask maintains high and nearly uniform contrast over extended distances, a key requirement for high-aspect-ratio lithography. (d) Scanning electron micrograph of a hexagonal close-packed (HCP) lattice with overall dimensions of 800 × 800 × 720~\textmu m$^3$. (e) Magnified view of the HCP lattice showing a wall width of $\sim$6~\textmu m, corresponding to an aspect ratio 120:1. (f) Because the entire structure is formed in a single exposure step with a typical duration of 20 s, the resulting lattice walls are smooth, in contrast to the rough surfaces commonly observed in structures fabricated by conventional two-photon polymerization lithography (see Supplementary Note S8).}\label{fig1}
\end{figure}

Here we introduce a single-exposure volumetric photolithography method that reconstructs a prescribed 3D intensity distribution using an inverse-designed phase mask (Fig.~\ref{fig1}). By explicitly constraining the optical field throughout an extended axial volume, the approach suppresses diffraction-induced contrast loss while preserving \textmu m-scale lateral resolution. This enables rapid fabrication of millimetre-scale structures with ultra-high aspect ratios in a single exposure, bridging the long-standing gap between high-resolution serial techniques and high-throughput volumetric manufacturing.

The engineered optical field exhibits an extended longitudinal profile while maintaining a lateral resolution of $\sim$4\textmu m. As a result, a uniform polymerization dose is delivered throughout the target volume of SU-8 photoresist (XFT-100, Kayaku), allowing high-aspect ratio microstructures, including hexagonal-cell lattices and MEMS-like structures, to be fabricated within a single $\sim$20 s exposure. In contrast, conventional photolithography is typically limited to patterning thin resist layers using binary photomasks. Under soft-contact exposure conditions, an air gap of 10 to 40~\textmu m is commonly present between the mask and the resist surface, leading to rapid diffraction-induced degradation of pattern fidelity with depth.

To quantify this limitation, we simulated light propagation from a chromium photomask containing a hollow hexagonal-cell aperture with an inner diameter of 80~\textmu m and a ring width of 4~\textmu m under 405 nm illumination (Fig. S1). The axial intensity distribution was computed using a scalar diffraction model, starting from a 20~\textmu m mask–resist gap and extending over a total propagation distance of 450~\textmu m in air (Fig.~\ref{fig1}a). While high contrast is maintained over the first several tens of micrometers, the pattern rapidly blurs beyond $\sim$80~\textmu m due to diffraction (Supplementary Note S1).

By comparison, an inverse-designed phase mask can strongly suppress diffraction-induced degradation, as previously demonstrated \cite{banerji2020extreme, lin2025diffraction}. As shown in Fig.~\ref{fig1}b, such a mask reconstructs high-contrast intensity patterns over an extended propagation distance within the photoresist. In this design, both the binary and inverse-designed masks target the fabrication of a hexagonal close-packed (HCP) lattice with a wall width of $\sim$4~\textmu m and a total depth of $\sim$720~\textmu m in SU-8 photoresist (refractive index 1.613 at 405 nm). Simulations in Fig.~\ref{fig1}c show that the contrast of the pattern produced by the binary mask degrades rapidly beyond $\sim$40~\textmu m, whereas the inverse-designed mask maintains high contrast $>$83\% across the full target depth (Fig. S1). The maximum achievable structure height is currently limited by the thickness of SU-8 that can be reliably deposited by spin coating, as discussed later.

Figures~\ref{fig1}(d–f) present scanning-electron micrographs at progressively higher magnification. These images first confirm the formation of the intended HCP lattice with an overall depth of $\sim$720~\textmu m, and subsequently resolve the individual hexagonal walls, which have a width of about 6~\textmu m, corresponding to an aspect ratio of 120:1. Notably, the entire structure is fabricated in a single exposure lasting $\sim$20 s. As a result, unlike voxel-based serial techniques such as two-photon polymerization lithography, the fabricated microstructures exhibit smooth and uniform walls, without stitching artefacts or surface roughness (Supplementary Note S8).

\section{Experimental methods}

\begin{figure}[h]
\centering
\includegraphics[width=1\textwidth]{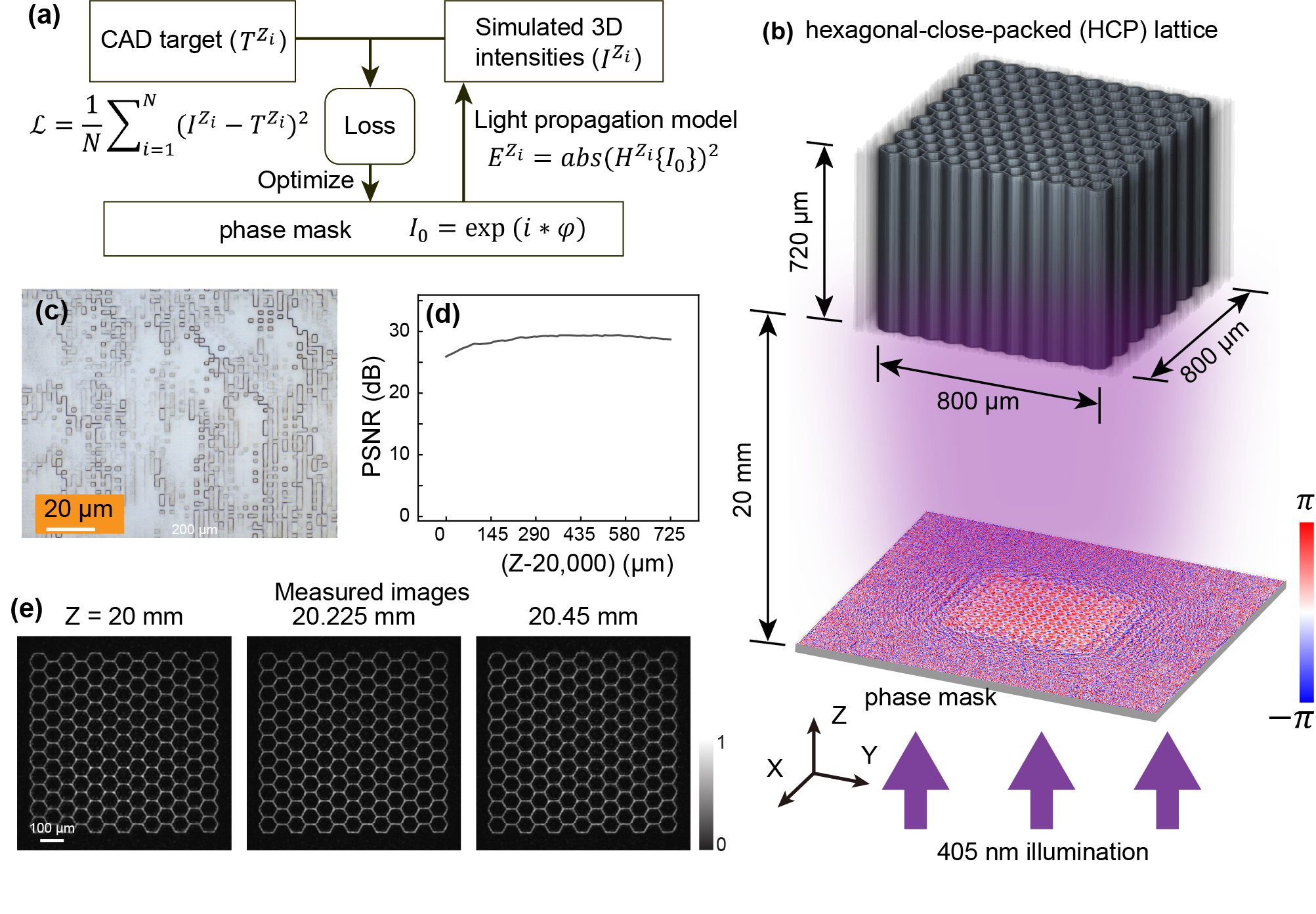}
\caption{\textbf{Inverse-design and optical characterization of phase mask.} (a) Schematic of the inverse-design workflow used to compute the phase mask that reconstructs a prescribed 3D intensity distribution. (b) The inverse-designed phase mask, shown by its phase profile (bottom), illuminated by a collimated laser beam at $\lambda=$ 405 nm to generate an extended hexagonal close-packed (HCP) lattice intensity pattern. (c) Optical micrograph of a representative region of the fabricated phase mask, confirming faithful realization of the designed phase features. (d) Peak signal-to-noise ratio (PSNR) extracted from the measured images as a function of propagation distance, indicating high image fidelity throughout the target volume. (e) Measured intensity images at different axial distances from the phase mask, corresponding to the mask–resist gap and propagation distances of Z = 20, 20.225 and 20.450 mm, demonstrating uniform image formation over an extended axial range. This range corresponds to effective propagation distances of 20, 20.36, and 20.72 mm in SU-8 photoresist (refractive index, n=1.613 at $\lambda=$~405 nm). Also see Fig. S2 and Supplementary Video 1. }\label{fig2}
\end{figure}

Volumetric exposure fields were generated using inverse-designed phase masks, as schematically illustrated in Fig.~\ref{fig2}a. The target 3D intensity distributions were specified from computer-aided design (CAD) models discretized into a series of axial slices. To compute the corresponding phase masks, we implemented a fully differentiable forward model in TensorFlow, enabling gradient-based optimization of the mask profile.

Forward propagation from the phase mask to the target volume was computed using the angular spectrum method, explicitly accounting for free-space propagation and the refractive-index mismatch between air and the photoresist. The phase profile was optimized by minimizing the mean-squared error between the reconstructed and target intensity distributions across all axial planes. This loss function enforces simultaneous fidelity throughout the full axial extent of the target volume, rather than at a single focal plane. Gradient-based optimization was used to iteratively update the phase profile until convergence (Supplementary Note S4) \cite{lin2026single, lin2025diffraction}. The optimized phase masks comprised $1000 \times 1000$ pixels with a pixel pitch of 2~\textmu m. As illustrated by the HCP lattice example in Fig.~\ref{fig2}b, the resulting designs maintain lateral confinement while strongly suppressing diffraction-induced degradation over extended axial distances.

The optimized phase masks were fabricated using grayscale optical lithography on a DWL~66+ direct-write lithography system (Heidelberg Instruments; Supplementary Note S5) \cite{menon2023perspectives}. A photoresist thickness matched to the designed phase depth was patterned to encode the continuous phase modulation. The fabricated masks were characterized using confocal optical profilometry to verify height fidelity and phase accuracy. An optical micrograph of a representative region of a fabricated phase mask is shown in Fig.~\ref{fig2}c. The measured height profiles closely match the design (Supplementary Note S5).

Optical characterization of the fabricated phase masks was performed by illuminating them with a collimated laser beam at $\lambda = 405$~nm, as shown in Fig.~\ref{fig2}b. The reconstructed 3D intensity distribution was relayed with unit magnification onto a CMOS image sensor. Axial scans were acquired by translating the sensor along the optical axis while maintaining conjugate-plane imaging. The measured intensity distributions were compared with simulations to assess image fidelity, contrast, and invariance with propagation distance. The resulting peak signal-to-noise ratio (PSNR) for the HCP lattice is plotted in Fig.~\ref{fig2}d, with representative reconstructed images shown in Fig.~\ref{fig2}e.

\begin{figure}[htb!]
\centering
\includegraphics[width=1\textwidth]{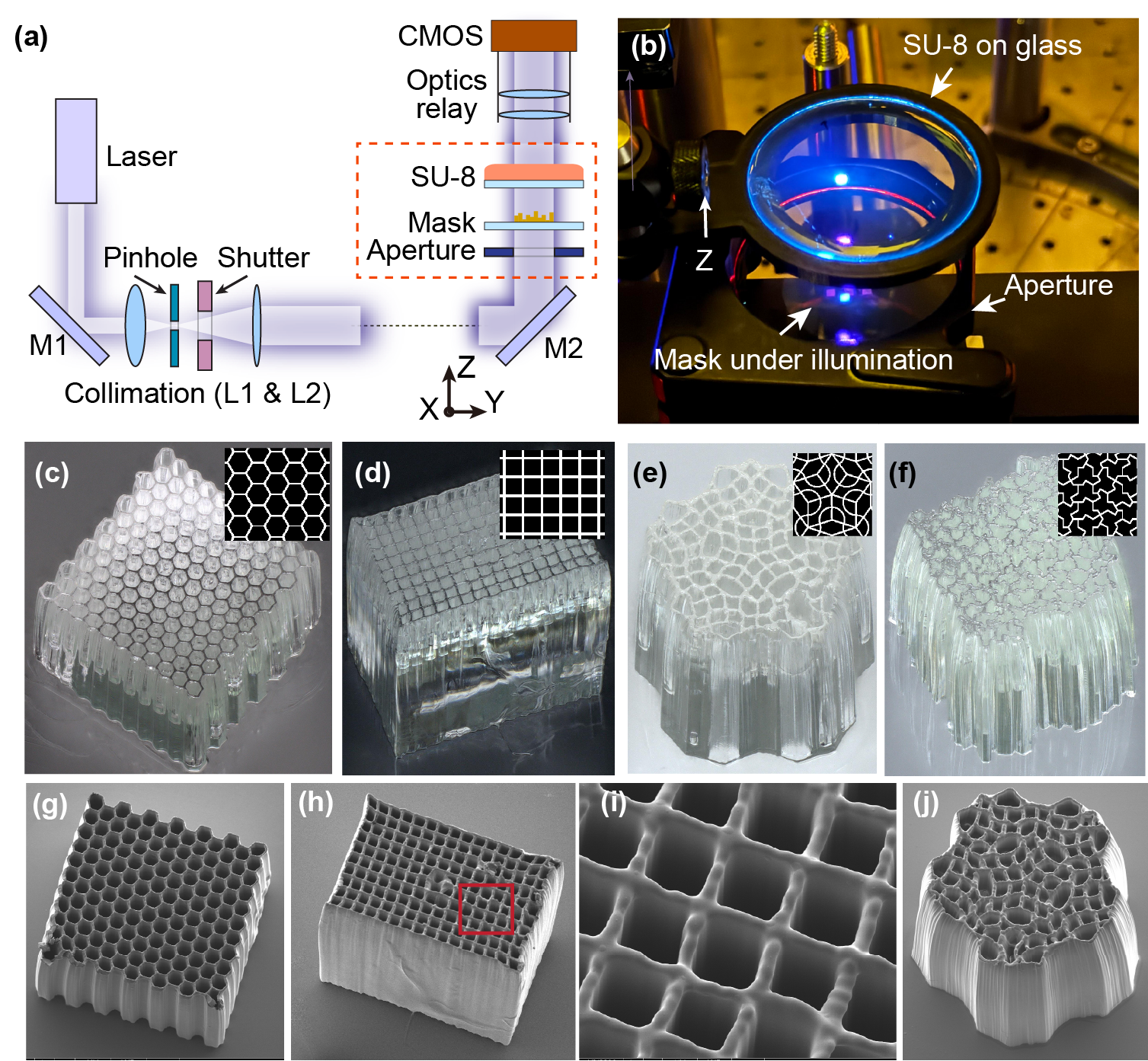}
\caption{\textbf{Single-exposure fabrication of high-aspect-ratio 3D microstructures.} (a) Schematic of the optical setup used for single-exposure volumetric lithography.(b) Photograph of the phase mask under illumination, corresponding to the region indicated by red dashed lines in (a). (c–f) Optical micrographs of developed SU-8 photoresist showing representative 3D geometries fabricated using a single exposure, including a hexagonal close-packed (HCP) lattice, a Cartesian lattice, a Penrose tiling, and the aperiodic “hat” monotile \cite{monotile}. All structures have nominal dimensions of 800 × 800 × 720 ~\textmu m$^3$ and were exposed for 20 s.  The target slices are indicated in the top-right inset in each panel. (g,h) Scanning electron micrographs of the HCP and Cartesian lattice structures, respectively.(i) Magnified view of the region highlighted by the red square in (h), resolving individual lattice walls. (j) Scanning electron micrograph of the Penrose tiling structure. All SU-8 structures were coated with a thin Ti/Pt layer to mitigate charging during electron microscopy. Also see Supplementary Note S9.}\label{fig3}
\end{figure}

To test single-exposure lithography, high-viscosity SU-8 photoresist was spin-coated onto 2-inch soda-lime glass wafers at 500 rpm for 60 s with a ramp rate of 200 rpm/s, yielding a resist thickness of $\sim$720~\textmu m. Soft baking was performed at 65~$^\circ$C for 5 min followed by 110~$^{\circ}$C for 80 min to fully solidify the resist and remove residual solvent. High-aspect-ratio microstructures were created by first exposing the SU8-coated glass wafers in a custom-built single-exposure lithography system (Figs.~\ref{fig3}a,b, and Supplementary Note S6). A 405-nm diode laser (Edmund Optics, \#19-462) was spatially filtered and collimated to provide uniform illumination of the hologram mask. The modulated optical field reconstructed the target 3D intensity distribution within the SU-8 volume after free-space propagation across a 20-mm air gap. 

Exposure times ranged from 12 to 32 s, depending on the target geometry and the desired wall thickness, with most structures fabricated using a nominal exposure time of 20 s (Supplementary Note S12). The exposure duration was precisely controlled using a computer-controlled shutter. Prior to exposure, the phase mask and SU-8-coated substrate were aligned along all three translational axes to ensure accurate registration (Supplementary Note S7).

Following exposure, samples were post-exposure baked at 65~$^\circ$C  for 5 min on a hot plate, followed by baking at 110~$^\circ$C for 120 min in a convection oven to promote uniform crosslinking throughout the full resist thickness. The samples were then developed in propylene glycol methyl ether acetate (PGMEA) for 60 min on an orbital shaker at 80 rpm, with extended development used as needed to ensure complete removal of unpolymerized resist from deep features. Finally, the samples were rinsed in isopropanol and dried under a gentle nitrogen stream. Optical micrographs of representative 3D structures, including a HCP lattice, a Cartesian lattice, a Penrose tiling, and the aperiodic “hat” monotile \cite{monotile}, are shown in Figs.~\ref{fig3}c–f. All structures have nominal dimensions of 800 × 800 × 720~\textmu m$^3$ and were fabricated using a 20 s exposure.

PEB introduces a small but reproducible asymmetry in feature dimensions, with slightly greater shrinkage observed at the surface exposed to the oven atmosphere than at the surface in contact with the glass substrate, consistent with the presence of a thermal gradient. In all cases, the resulting wall thickness was $\sim$6~\textmu m.

Lastly, the structures were characterized using a field-emission scanning electron microscope (FEI Quanta 600F). Prior to imaging, samples were sputter-coated with a 10-nm Ti adhesion layer followed by a 50-nm Pt layer to minimize charging. Imaging was performed at 10 kV accelerating voltage with working distances of $\sim$18 mm. Tilted views were acquired to assess sidewall verticality, surface morphology, and structural integrity of high-aspect-ratio features. Exemplary micrographs are shown in Figs.~\ref{fig3}(g-j).

\section{Results}
\begin{figure}[htb!]
\centering
\includegraphics[width=0.7\textwidth]{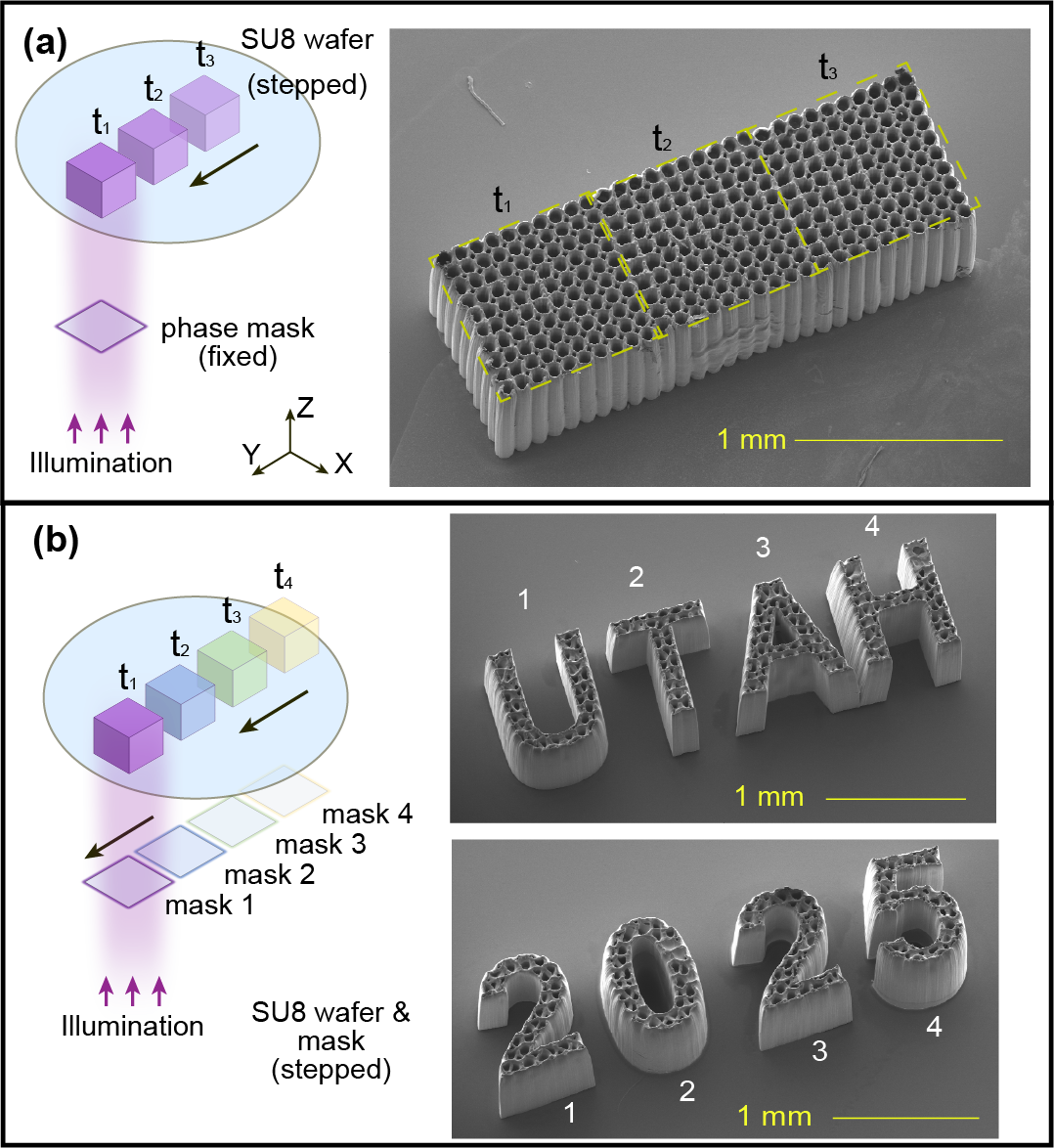}
\caption{\textbf{Enhancing complexity of printed microstructures via stepped exposures.} (a) By stepping the SU-8 wafer relative to the mask and doing multiple exposures, one can tile the HCP lattice to create a much larger HCP lattice. The phase mask is kept fixed in this example. (b) By stepping the SU-8 wafer and also using different masks, one can combine multiple patterns to create a larger structure.  Additional images including prints from single exposures are provided in Fig. S7.}\label{fig4}
\end{figure}

Our single-exposure lithography platform is inherently scalable and well suited for large-area, high-throughput fabrication. As illustrated in Fig.~\ref{fig4}, both the SU-8–coated substrate and the holographic phase mask can be mounted on motorized translation stages to enable automated, sequential patterning. In one operating mode (Fig.~\ref{fig4}a), the phase mask remains fixed while the substrate is translated between exposures, allowing rapid replication of identical structures. Using this configuration, we fabricated a tiled circular-cell lattice comprising three adjacent units with overall dimensions of 2400 × 800 × 720~\textmu m$^3$ and wall thickness of 6~\textmu m. The structure was completed in 68 s, consisting of three 20 s exposures interleaved with two 4 s translation and stabilization steps, yielding a threefold increase in fabrication volume relative to a single exposure. This approach further enables the fabrication of spatially separated arrays across larger substrates, with each structure produced in $\sim$20 s. At this rate, a production throughput exceeding 4,000 structures per day is readily achievable.

In a second configuration, both the phase mask and the substrate are synchronously translated, enabling sequential patterning with multiple, independently designed masks (Fig.~\ref{fig4}b). Using this scheme, we fabricated stacked 3D lettering spelling “UTAH” followed by “2025”, with each character defined by a distinct mask. All features employed 6~\textmu m wall thicknesses and incorporated a triangular infill for mechanical stability. The resulting four-layer structure has overall dimensions of approximately 3000 × 800 × 720~\textmu m$^3$ and was completed with a total exposure time of 92 s, including four 20 s exposures and three 4 s translation intervals. This configuration highlights the flexibility of the platform for customized, multi-pattern volumetric fabrication.

Because holographic reconstructions exhibit a nonzero background intensity arising from zero-order light and speckle, fabricating multiple structures within a shared exposure field can lead to cumulative background exposure and reduced contrast between target and non-target regions. As a result, structures that can be reliably fabricated with exposure times of 20–60 s when produced individually must be exposed for shorter durations when patterned in arrays. In practice, exposure times of 20 to 40 s per structure were optimal, whereas longer exposures resulted in unintended curing within interior regions. We also note that advanced proximity-effect correction can be readily employed to these structures, if required.\cite{wan2015proximity} 

\begin{figure}[htb!]
\centering
\includegraphics[width=0.81\textwidth]{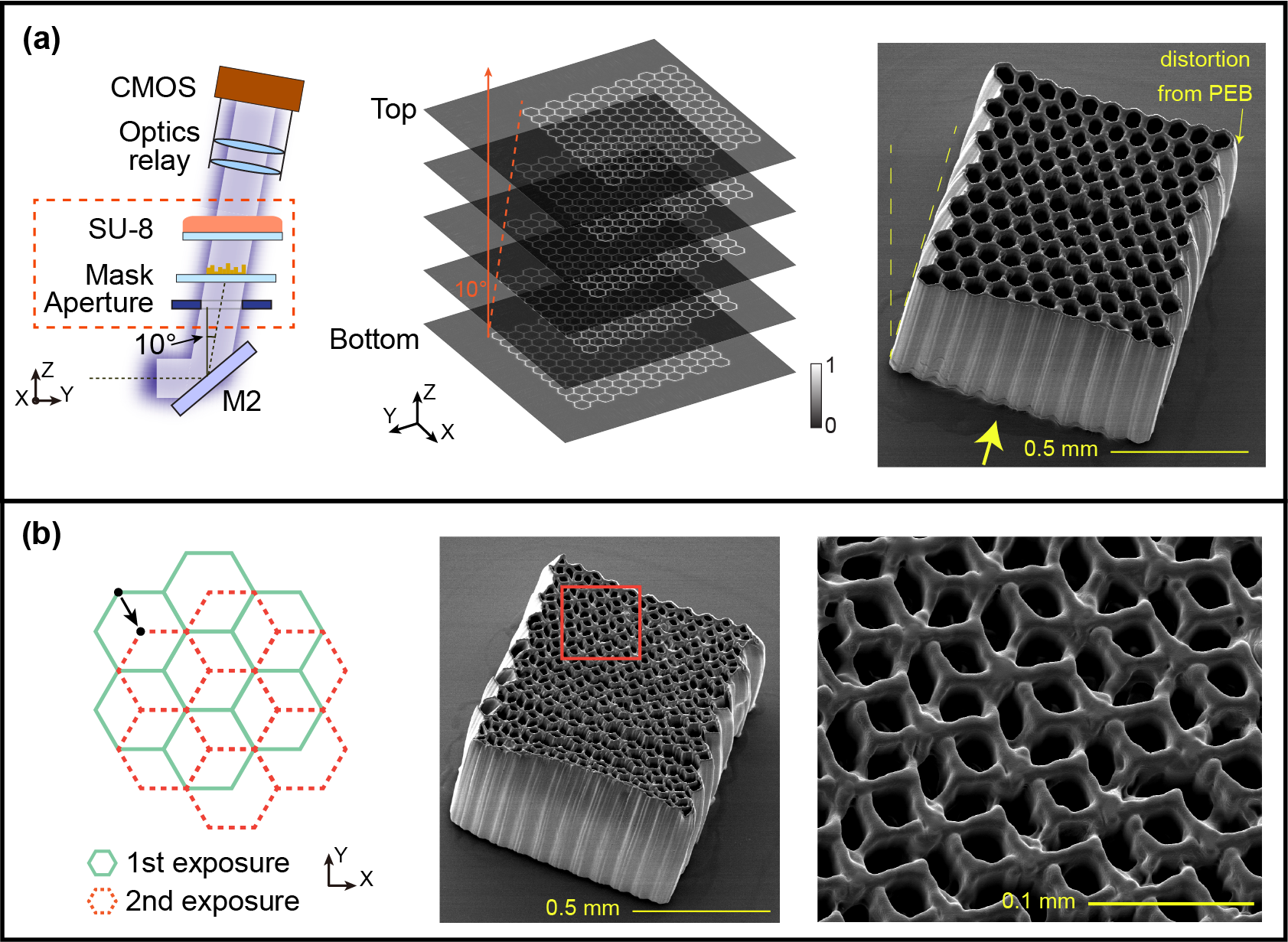}
\caption{\textbf{Enhancing complexity of printed microstructures via tilted and overlapping exposures.} (a) By tilting the illumination relative to the mask, we create tilted 3D intensity distribution inside the SU-8, thereby creating a tilted HCP lattice. The monitoring optics, including the CMOS sensor and relay optics, are tilted to maintain mask–resist alignment. The middle panel illustrates recorded images as function of distance from the mask illustrating the tilt in the 3D pattern. (b) Complex structures can be created by overlapping two exposures with a relative translation between them as illustrated in the left panel. composite structures. In the scanning-electron micrograph, we illustrate double exposure with a HCP lattice, and the magnified views confirm the increased geometric complexity. Additional images including prints from single exposures are provided in Fig. S8.}\label{fig5}
\end{figure}
To demonstrate the versatility of the single-exposure platform for fabricating high-aspect-ratio 3D structures, we consider two representative strategies: oblique illumination and overlapping exposures, as illustrated in Fig.~\ref{fig5}. Rather than conventional normal-incidence illumination, a 10$^\circ$ oblique beam is used to intentionally tilt the reconstructed volumetric intensity distribution (Fig.~\ref{fig5}a). The additional linear phase imposed by oblique illumination results in a systematic lateral shift of the reconstructed field with propagation distance. Consistent with this expectation, the measured intensity distribution (middle panel in Fig.~\ref{fig5}a) exhibits a lateral displacement of approximately 80~\textmu m between the bottom and top planes over a propagation distance of 450~\textmu m in air, in good agreement with the predicted shift, $\Delta y = z \tan \theta$. The corresponding tilted HCP lattice, fabricated with an exposure time of $\sim$20 s, is shown in right panel of Fig.~\ref{fig5}.

We further demonstrate a double-exposure strategy to generate more complex 3D architectures by sequentially overlapping independently designed intensity fields. Between exposures, the SU-8 substrate was laterally translated by 20~\textmu m along $x$ and 34~\textmu m along $y$ (Fig.~\ref{fig5}b). The resulting composite HCP lattice, fabricated using two 15 s exposures, is shown in the right panel in Fig.~\ref{fig5}b, and highlights the increased pattern density and structural complexity enabled by this approach.

\begin{figure}[htb!]
\centering
\includegraphics[width=\textwidth]{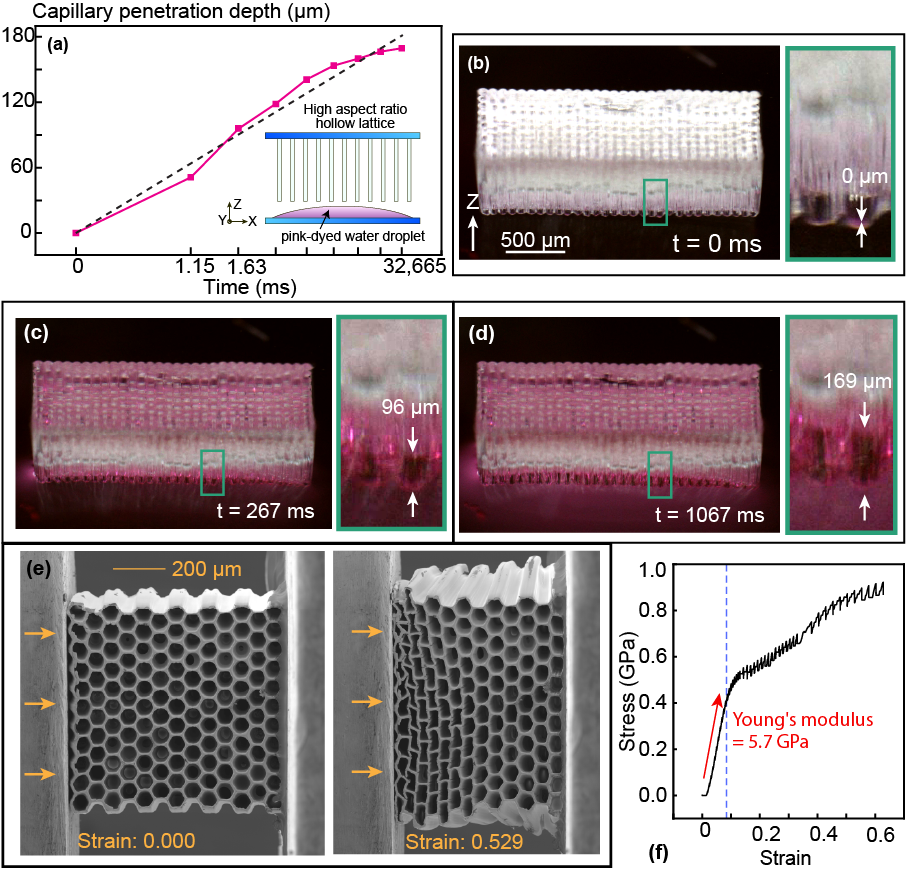}
\caption{\textbf{Capillary flow through high-aspect ratio hollow lattice and Nano-indentation experiments.} (a) Measured penetration of water into a hollow circular lattice as a function of time, exhibiting the characteristic square-root dependence expected for capillary-driven flow. The inset shows a schematic of the experiment, illustrating a dyed water droplet poised to enter the hollow SU-8 lattice. (b-d) Time-resolved photographs of the device at 0, 267, and 1,067 ms, respectively, with magnified views shown in the right panels (Supplementary Video 2). (e) Time-resolved scanning-electron micrographs at compressive strain = (left) 0 and (right) 0.529 captured during a nano-indentation experiment (Supplementary Video 3). Arrows indicate regions of localized bending and densification, as well as direction of motion of the indenter. (f) Corresponding stress–strain curve derived from load–displacement data, with Young’s modulus (0.59 GPa) extracted from the linear elastic regime (dashed line). Also see Supplementary Note S15.}\label{fig:cap}
\end{figure}
Capillary flow behavior was characterized using circular-cell lattice structures (inner channel diameter of 68~\textmu m and a depth of 720~\textmu m) prior to metal deposition. The sample was inverted, mounted on a wafer holder, and positioned with sub-micrometre precision using a three-axis translation stage to bring the structure into controlled contact with the liquid surface as illustrated in the inset in Fig~\ref{fig:cap}a. Experiments were performed prior to metal coating, and the liquid was dyed to enhance optical contrast. 

Capillary infiltration was monitored by optical microscopy (Keyence VHX-X1) at a 45$^\circ$ viewing angle. Upon contact with water, rapid capillary-driven imbibition was observed, with the penetration depth reaching approximately 169~\textmu m within the first second. The infiltration rate subsequently decreased, reaching 233~\textmu m at 10 s and saturating at approximately 243~\textmu m by 20 s, consistent with decelerating Washburn-type capillary dynamics.\cite{courbin2007imbibition, park2016capillarity} The limited ultimate penetration depth arises from incomplete lattice development near the base of the structure, where residual photoresist partially obstructs the channels, an effect that can be mitigated by longer development times. Representative frames from Supplementary Video 2 are shown in Figs. \ref{fig:cap}(b–d).

The measured infiltration dynamics follow classical Lucas–Washburn behavior, in which the penetration depth L scales as L $\propto \sqrt{\text{t}}$ due to the balance between capillary pressure and viscous dissipation. Fitting the early-time data (t $\leq$ 10 s) to a square-root dependence yields good agreement with this model, confirming that liquid transport within the hollow lattice is governed primarily by capillary forces rather than inertial or gravitational effects. Deviations from ideal Washburn scaling at longer times are attributed to increased hydraulic resistance caused by partial channel obstruction near the base of the structure. The fit, shown as a dashed line in Fig. \ref{fig:cap}a, yields an effective capillary transport coefficient of 176.3~\textmu m/$\sqrt{\text{ms}}$. These results demonstrate that the high-aspect-ratio tubular lattices support efficient capillary-driven transport, highlighting their potential for microfluidic and bio-integrated applications in which fluid guidance is achieved through surface tension and adhesion, without the need for external pressure.

The mechanical response of the printed HCP lattice was quantified by in situ uniaxial micro-compression using a flat-punch nanoindenter (\textmu TR, MTR-3-XR) inside a scanning electron microscope (FEI TENEO, Supplementary Note S16). The structure exhibits a well-defined linear elastic regime at small strains ($\epsilon < 9$\%), corresponding to an effective Young’s modulus of $\sim$5.7~GPa, with deformation dominated by uniform elastic bending of the cell walls. Beyond this regime, the stress–strain response becomes nonlinear due to the onset of localized buckling and progressive cell-wall collapse, producing a characteristic serrated response associated with sequential failure of lattice columns. At larger strains, the structure undergoes crushing and densification, accompanied by partial elastic recovery upon unloading in regions that remain intact. These results confirm that the single-exposure printed architectures possess substantial stiffness and mechanically robust lattice behavior despite their high aspect ratio and complex 3D geometry.

\section{Discussion}
To establish fundamental limits on achievable aspect ratio and pattern density in single-exposure volumetric photolithography, we numerically investigated HCP lattices while systematically increasing the target structure depth from 0 to 6,400~\textmu m in 800~\textmu m increments, with the hexagonal cell diameter fixed at 80~\textmu m (Supplementary Note S13). In cross section, each hexagonal ring exhibits a Gaussian-like lateral intensity profile, and the effective feature width is quantified by the full width at half maximum (FWHM). For a zero-depth target corresponding to a single-layer pattern, the minimum FWHM is $\sim$6~\textmu m, approaching the diffraction-limited resolution of 4~\textmu m set by the system numerical aperture (NA = 0.05).

As the designed axial extent increases, lateral feature widths broaden monotonically, reaching $\sim$16~\textmu m at a depth of 6,400~\textmu m, corresponding to a 2.5-fold increase relative to the diffraction-limited value. This behaviour is consistent with wave-optical constraints associated with extended axial propagation of engineered interference fields, analogous to quasi–diffraction-free Bessel-like beams formed through conical wave superposition. Over long propagation distances, energy redistribution from the central lobe into higher-order side lobes increases overlap within individual unit cells, leading to progressive degradation of optical contrast (Supplementary Note S13). Importantly, this broadening reflects a fundamental consequence of volumetric wavefront encoding rather than a limitation of the inverse-design algorithm.

Optical contrast provides a complementary metric for assessing printability. The single-layer pattern exhibits the highest contrast (92\%), whereas increasing target depth leads to a rapid reduction in contrast beyond the designed axial region. At sufficiently large depths, the contrast can become negative, with partial recovery attributable to Talbot-type self-imaging effects. However, the depth-averaged contrast decreases monotonically with increasing structure height, reaching $\sim$50\% at 6,400~\textmu m. This contrast is likely insufficient to ensure reliable polymerization across the full depth, thereby setting a practical constraint on the achievable aspect ratio in single-exposure volumetric patterning.

We further examined the influence of pattern density by varying the hexagonal cell diameter from 40 to 160~\textmu m at a fixed target depth of 3,200~\textmu m (Supplementary Note S13). Larger unit cells consistently exhibit higher optical contrast, owing to reduced side-lobe overlap and weaker inter-cell interference. These results reveal an intrinsic trade-off between lateral feature density and achievable depth, indicating that lower-density patterns are inherently more robust for extreme aspect-ratio fabrication. Although quantified here for HCP lattices, this trade-off is expected to apply broadly to volumetric interference-based fabrication strategies that rely on extended axial coherence.

Related approaches for engineering 3D optical intensity distributions have previously been explored in multi-plane lithography and projection-based methods\cite{wang2015optical, menon2020methods, menon2015nanophotonic}, as well as in holographic field synthesis\cite{mohammad2017full, meem2019multi}. In contrast to plane-wise or projection-limited strategies, the present work explicitly engineers continuous volumetric intensity distributions and quantitatively identifies the physical trade-offs governing depth, resolution and contrast. These results provide a pathway to expand the accessible design space for scalable, high-fidelity volumetric micro- and nanofabrication.

\section{Conclusion}
Here we demonstrate a single-exposure volumetric photolithography approach that relaxes the conventional trade-off between resolution, aspect ratio and throughput in 3D microfabrication. By inverse-designing a phase mask to reconstruct a prescribed 3D dose distribution\cite{yu2023ultrahigh, dorrah2023light, makey2019breaking, sun2022three}, this method sustains usable optical contrast over mm-scale depths while maintaining \textmu m-scale lateral feature sizes. As a result, mm-scale architectures with sub--10~\textmu m features and aspect ratios exceeding 120:1 can be fabricated in a single 20~s exposure, without layer-by-layer stacking, voxel-wise scanning or interlayer registration.

Beyond enabling ultra-high-aspect-ratio fabrication, this work establishes inverse-designed wavefront engineering as a general framework for deterministic volumetric patterning in thick photoresists. Encoding complex intensity distributions enables the fabrication of periodic and aperiodic architected lattices with smooth sidewalls and high structural fidelity. The absence of layer interfaces yields mechanically continuous architectures with enhanced robustness, as confirmed by in situ micro-compression measurements, while hollow lattice geometries support controlled capillary-driven fluid transport consistent with classical scaling laws.

The platform is inherently scalable and reconfigurable. Stepped, tilted and overlapping exposures enable rapid tiling, multi-pattern integration and controlled geometric distortion, allowing complex quasi-3D architectures to be assembled without sacrificing throughput. Quantitative analysis further reveals fundamental design trade-offs between pattern density, achievable depth and optical contrast, providing clear guidelines for extending this approach toward larger volumes and higher structural complexity.

Together, these results provide a scalable route to high-fidelity volumetric fabrication that bridges the gap between high-resolution serial nanofabrication and high-throughput manufacturing. By combining inverse-designed optics with single-exposure lithography, this work defines a new regime of volumetric fabrication that is compatible with established photolithographic infrastructure and broadly applicable to architected materials, microfluidics, MEMS and micro-optical systems.
\newpage
\backmatter
\bmhead{Supplementary information}
Additional information is available in the Supplementary document and Supplementary Videos.

\bmhead{Acknowledgements}
We thank Chi-Hao Chang, Kwong Sang Lee and Michael Cullinan for helpful discussions; Joseph Jacob and Brian van Devener for fabrication support; and Bob Wheeler for nanoindentation support. This work made use of the Nanofab EMSAL shared facilities of the Micron Technology Foundation Inc. Microscopy Suite and was performed in part at the Utah Nanofab Cleanroom, supported by the John and Marcia Price College of Engineering, the Health Sciences Center, and the Office of the Vice President for Research. The University of Utah Nanofab shared facilities are supported in part by the National Science Foundation MRSEC Program under Award No. DMR-112125.

\begin{itemize}
\item Funding\\
US National Science Foundation Future Manufacturing grant \#2229036.\\
US DARPA HR0011-25-9-0019.

\item Conflict of interest\\
RM has financial stake in Oblate Optics, Inc. which is commercializing inverse-designed flat optics. RM is a co-inventor of a related patent filed by University of Utah and the University of Texas at Austin, Serial No. WO202525956. RM is a co-inventor of a related granted patent (2015) US 8,953,239. 

\item Data availability\\ 
All data is available from the corresponding author upon reasonable request. 

\item Author contribution\\
DL: Inverse design; Mask fabrication; Resin formulation and characterization; 3D printing system
setup; 3D printing experiments; Acquisition, analysis, or interpretation of data; Revising and editing
the manuscript critically for intellectual content.\\
BB: Fabrication and metrology of the phase mask.\\ 
RM: Conception or design of the study; Supervision; Funding acquisition; Revising and editing the
manuscript critically for intellectual content.
\end{itemize}


\bibliography{sn-bibliography}

\end{document}